\documentclass[reprint,twocolumn,amsmath,amssymb]{revtex4}
\usepackage[utf8]{inputenc}
\usepackage[T1]{fontenc}
\usepackage[english]{babel}
\usepackage{graphicx}
\usepackage{hyperref}
\hypersetup{citecolor=black, linkcolor=black, urlcolor=black, colorlinks=true}

\newcommand{\bn}{\mathbf{n}}

\newcommand{\id}{\mathrm{d}}
\newcommand{\br}{\mathbf{r}}
\newcommand{\bR}{\mathbf{R}}
\newcommand{\bG}{\mathbf{G}}
\newcommand{\bg}{\mathbf{g}}

\newcommand{\bfF}{\mathbf{F}}

\begin{document}

\title{Switching of Nonequilibrium Depletion Force Caused by Blockade Effect}

\author{S.P.~Lukyanets} \email[Email address:
]{lukyan@iop.kiev.ua}

\author{O.V.~Kliushnychenko}
\affiliation{Institute of Physics, National Academy of Sciences,
Prospect Nauky 46, 03028 Kyiv, Ukraine}

\begin{abstract}
The concentration-dependent switching of the non-equilibrium depletion forces between obstacles in an interacting Brownian gas flow is presented. It is shown that this switching is caused by the blockade effect for the gas particles. With increasing equilibrium gas concentration, the gas particles blockade causes the obstacle wake inversion (trace profile ``turn-over'') that, in turn, leads to the change of sign of dissipative interaction. Some non-linear effects such as formation of a cavity-like sparse wake behind the obstacle and the dissipative pairing effect are discussed briefly. The results are obtained within the lattice gas model in the mean-field approximation.
\end{abstract}

\maketitle
\section{Introduction}\label{intro}
The motion of impurities in a Brownian gas is accompanied by its density perturbation (impurity trace, or wake) that, in turn, induces additional non-equilibrium correlations between the impurities. These correlations, called dissipative or non-equilibrium depletion forces \cite{dzubiella_depletion_2003,Sasa_2006,khair_motion_2007,sriram_out--equilibrium_2012}, are responsible for the coherent part of the collective friction force as well as for possible dissipative structures of impurity ensemble.

In equilibrium, the depletion interaction is usually short-range; the spatial range is of the order of the depletion agent characteristic length scale \cite{lekkerkerker_2011,crocker_entropic_1999}. In contrast, the non-equilibrium forces between the impurities may exhibit long-range behavior due to a long-living diffusion trace induced by their motion \cite{Dean2010,demery_2014,benichou_stokes_2000,benichou_force_2001,benichou_biased_2013}.
In addition, such forces often have unusual properties, e.g., they violate the Newton's third law \cite{dzubiella_depletion_2003,Sasa_2006,pinheiro_2011,Ivlev_2014}.

It is clear that specific properties of the dissipative interaction are determined by the structure and the shape of the perturbation profile, which, in turn, significantly depends on the interaction between the gas particles.

The presence of inter-particle interaction, even short-range one (as in a lattice gas, when a lattice site  can be occupied by only one particle), is responsible for a number of unexpected kinetic effects, e.g., ``back correlations'' effect \cite{tahir-kheli_correlated_1983}, drifting spatial structures \cite{schmittmann_statistical_1995,leung_drifting_1997,hipolito_effects_2003}, effects of ``nega\-tive'' mass transport \cite{Lukyanets2010,argyrakis_negative_2009,efros_negative_2008}, dissipative pairing effect for tracers passing through a lattice gas \cite{mejia-monasterio_bias-_2011}.

An important mechanism often underlying such kinetic phenomena is a blockade effect. As was shown in \cite{kliushnychenko_blockade_2014}, the blockade effect implicates significant changes in the form of an obstacle trace. The trace structures for the cases of impurity moving in a lattice gas and fixed impurity or obstacle in a gas flow do not coincide in general. In particular, the obstacle trace can take atypical form characterized by switching of its direction depending on the equilibrium gas concentration. Usually, the profile of the gas perturbation induced by moving inclusion has a dense localized region ahead of the inclusion and a long low-density tail behind it, which damps according to a power law, e.g., \cite{Dean2010,demery_2014,benichou_stokes_2000,benichou_force_2001,benichou_biased_2013}. However, for an obstacle in a gas flow, the gas perturbation profile can take reversed structure with an extended dense region ahead of the obstacle and a localized low-density region in its wake \cite{kliushnychenko_blockade_2014}. It should be noted that similar effect in one-dimension was discussed in \cite{kolomeisky_asymmetric_1998}. This switching of the profile directivity  is caused by the  effect of particles blockade ahead of the obstacle. The blockade is due to hard-core-like interaction between the particles and is possible if the flowing gas density is high enough.

As the inversion of density perturbation takes place for arbitrary number and configuration of obstacles, see \cite{kliushnychenko_blockade_2014}, we may expect that this inversion has to affect the dissipative interaction between the obstacles.

\smallskip
The aim of this paper is to show that trace inversion is responsible for the change in sign of the induced non-equilibrium interaction between the obstacles, e.g., switching of the effective attraction to repulsion.

We consider the simplest model of a lattice gas with two obstacles whose sizes are much larger than the lattice constant. We use the mean-field and the long-wavelength approximations, neglecting short-range correlations and fluctuations in the gas, see \cite{kliushnychenko_blockade_2014}.

\section{Trace inversion in the lattice gas model}

As was shown in \cite{kliushnychenko_blockade_2014,Lukyanets2010}, the problem of an obstacle in a lattice gas flow can be considered as a limiting case of a two-component gas, one of the components is at rest while another one is flowing in a uniform external field.

We resort to the simplest model of a two-component lattice gas, when each lattice site can be occupied only by one particle, see \cite{tahir-kheli_correlated_1983}. The kinetics of multi-component lattice gas is defined by the jumps of particles to the neighboring vacant sites. The variation of the $i$-th site occupancy by the particles of $\alpha$ sort
during the time interval $\Delta t$, $\tau_0\ll\Delta t\ll\tau_l$ ($\tau_0$ is the duration of a particle jump to a neighboring site, $\tau_l$ being the lifetime of a particle on a site), is described by the standard continuity equation (see, e.g., \cite{chumak_diffusion_1980,tahir-kheli_correlated_1983})
\begin{equation}\label{balance+}
n_i^\alpha (t+\Delta t)-n_i^\alpha(t)=\sum_j
\left(J^\alpha_{ji}-J^\alpha_{ij}\right) + \delta J_i^\alpha,
\end{equation}
where $n_i^\alpha=0,1$ are the local occupation numbers of particles $\alpha$ at the $i$-th site, $J^\alpha_{ij}=\nu^\alpha_{ij}n_i^\alpha \left(1-\sum_\beta n_j^\beta\right) \Delta t$ gives the mean number of jumps (of $\alpha$ particles from site $i$ to a neighboring site $j$ per time $\Delta t$, $\beta(\alpha)$ indexes particle species), $\nu_{ij}^\alpha=\nu^\alpha$ is the mean frequency of these jumps. The term $\delta J_i^\alpha=\sum_j(\delta J_{ji}^\alpha-\delta J_{ij}^\alpha)$ stands for the Langevin source that is defined by the fluctuations $\delta J_{ji}^\alpha$ of the number of jumps between sites $j$ and $i$ during $\Delta t$ \cite{chumak_diffusion_1980}.
These fluctuations are caused by the fast, as compared to the time
scale $\Delta t$, processes and will be neglected for simplicity. It means that we neglect fluctuation-induced forces that may be appreciable for closely located inclusions.

In what follows we consider only two components, denoted by $n$ and $v$. In the absence of external fields for the regular lattice we suppose that $\nu_{ji}^n=\nu=\mathrm{const}$ for the mobile component $n$, while the component $v$ is assumed to be at rest, $\nu_{ji}^v=0$. The presence
of driving field leads to asymmetry of particle jumps.
Supposing the activation mechanism of jumps and the driving field
$\bG$ to be weak, the frequency may be written as
$\nu_{ji}^n\approx\nu[1+(\bG,\br_i-\br_j)/(2kT)]$, or
$\nu^\pm\approx\nu\pm\delta\nu$, where $\nu^+$ denotes the jump
frequency along the field, $\nu^-$ \--- against it, $\delta\nu=\nu
a|\bG|/(2kT)$ ($a$ is the lattice constant), condition
$a|\bG|/(2kT)<1$ being satisfied.

Equations for the average local occupation numbers can be obtained
from Eqs.~(\ref{balance+}) using the local equilibrium
approximation (Zubarev approach)
\cite{chumak_diffusion_1980,zubarev_nonequilibrium_1974}
which coincides, in our case,  with the mean field approximation
\cite{leung_novel_1994}. Introducing time derivatives
\cite{richards_theory_1977}, in the long-wavelength approximation,
see
\cite{schmittmann_statistical_1995,leung_drifting_1997,hipolito_effects_2003,Lukyanets2010},
macroscopic kinetics of the mobile component $n$ is given by the
equation
\begin{equation}\label{meqd}
  \partial_\tau n = \nabla^2n-\nabla(v\nabla n - n\nabla v)-(\bg,\nabla)[n(1-v-n)],
\end{equation}
where $n=n(\br,\tau)$ and $v=v(\br)$ are the average occupation
numbers of the two components at the point $\br$,
$\bg=a\bG/(2kT)$. Here, we have introduced dimensionless spatial coordinate
$\br/a\rightarrow\br$ and time $\tau=\nu t$, and $\partial_\tau$ stands for the partial time derivative.

Second term on the right-hand side of Eq.~(\ref{meqd}) is a reduced nonlinear mixing flow which describes mutual drag of one component by another and arises as a result of the particle distinguishability and the local repulsion (due to excluded volume constraint) in a lattice gas \cite{Lukyanets2010}.
This flow leads to the array of anomalous diffusive transport effects, e.g., the drag effect, formation of the drifting spatial structures \cite{schmittmann_statistical_1995,leung_drifting_1997,hipolito_effects_2003}, effects of ``nega\-tive'' mass transport \cite{Lukyanets2010,argyrakis_negative_2009,efros_negative_2008}, and induced long-time correlations \cite{kliushnychenko_induced_2013}.

\begin{figure}
\includegraphics[width=.98\columnwidth]{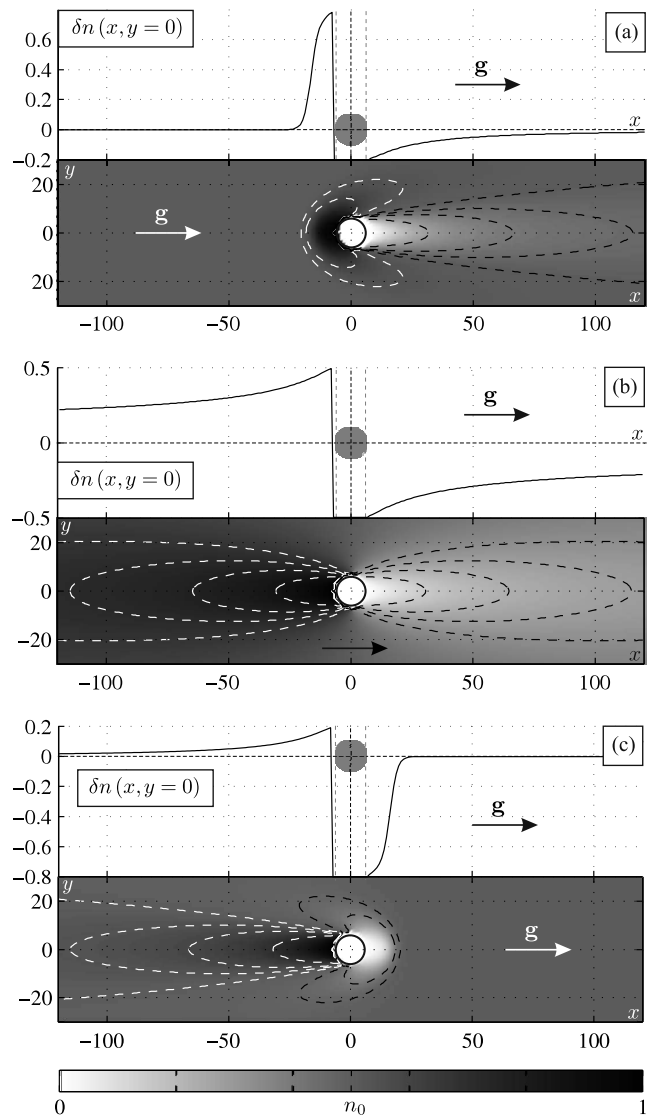}
\caption{\label{fig1} The two-dimensional equilibrated concentration distributions $\delta n(x,y)=n(x,y)-n_0$ (mean occupation numbers) of the lattice gas particles near the inclusion at the different values of equilibrium concentration $n_0$ are presented on contour plots: (a) $n_0=0.2$, (b) $n_0=0.5$, (c) $n_0=0.8$. The external field $\bg$ ($|\bg|=0.5$) is directed along the $x$-axis; the impermeable ($n_0=1$) circular inclusions of radius $R=7$ (in units of $a$) are placed at the origin (denoted by circles). The gray background corresponds to the equilibrium gas concentration $n_0$ for every contour plot, in consistence with the colorbar. Dashed contour lines underline the exact inversion between profiles (a) and (c), while the corresponding one-dimensional profiles $\delta n(x)=n(x,y=0)-n_0$ show the ``switching'' of asymptotic behavior.}
\end{figure}

\medskip
In order to demonstrate the trace inversion effect mentioned above we now consider a single macroscopic obstacle of a circular form in a flow of driven lattice gas particles. Fig.~\ref{fig1} presents the two-dimensional numerical solutions of Eq.~(\ref{balance+}) in the mean-field approximation for different equilibrium concentrations $n_0$.

For low concentrations ($n_0<1/2$), the structure of the profile is typical
\cite{dzubiella_depletion_2003,khair_motion_2007,sriram_out--equilibrium_2012,benichou_biased_2013} and is characterized by a lengthy depleted region behind the inclusion (wake) and a localized dense region in front of it, Fig.~\ref{fig1}a.

\begin{figure}
\includegraphics[width=.98\columnwidth]{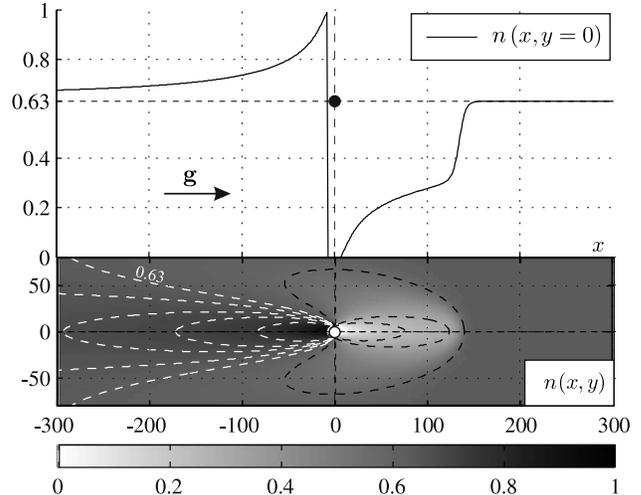}
\caption{\label{fig2}
The example stationary wake (numerical result) whose profile $m(x,y=0)$ is of the form of Bloch wall, describing the cavity behind the inclusion (black circle); equilibrium medium concentration $m_0=0.63$, field $|\bg|=0.5$, inclusion radius $R=7$.
}
\end{figure}

At high concentrations ($n_0>1/2$), the profile takes unconventional form with an extended dense region in front of the inclusion and a localized depleted one in its wake, resembling the form of a cavity, Fig.~\ref{fig1}c or Fig.~\ref{fig2}. Hence, the main part of the density perturbation is now shifted to the region in front of the obstacle and directed upstream.

This behavior is caused by the blockade effect of the gas particles flow due to the short-range repulsion between them (the condition that a lattice site can be occupied by only one particle). For sufficiently high concentrations $n_0$, gas particles have no time to leave the blockade zone ahead of the obstacle via lateral diffusion and, as a result, the dense region ahead of the obstacle has to grow. The blockade effect is non-linear and becomes significant near the obstacle surface and/or for large obstacles whose size is much larger than the lattice constant.

The trace inversion directly follows from the symmetry of Eq.~(\ref{meqd}) that can be represented in the form
\begin{equation}\label{meq}
      \partial_\tau n = \nabla(h\nabla n - n\nabla h -\bg n h),
\end{equation}
where $h=1-m-v$ is the vacancy concentration, i.e., concentration of the empty lattice sites. Equation (\ref{meq}) is invariant under transformation $n\leftrightarrow h$, $\bg\leftrightarrow-\bg$ and can easily be rewritten as
\begin{equation}\label{heq}
    \partial_\tau h = \nabla(n\nabla h - h\nabla n +\bg n h).
\end{equation}

Equation (\ref{meq}) describes the kinetics of the gas particles in sweeping field $\bg$ at equilibrium gas concentration $n_0$ while Eq.~(\ref{heq}) corresponds to the vacancy kinetics in the opposite field $-\bg$ at the equilibrium concentration of vacancies $h_0=1-n_0$. The tail of the obstacle trace in the gas flow is characterized by a depleted gas region behind the obstacle, whereas the trace in the corresponding vacancy flow (which has opposite direction) has a depleted tail of vacancies that corresponds to a dense region of gas particles.

For distribution $v(\br)$ with mirror symmetry along the field direction $\br_\parallel\parallel\bg$, i.e., $v(\br_\parallel)=v(-\br_\parallel)$, each density perturbation profile $\delta n(\br,n_0)=n(\br,n_0)-n_0$ at a given concentration $n_0$ can be obtained by the inversion transformation of the profile $\delta n(\br,1-n_0)$ at concentration $1-n_0$, that is expressed by relation
\begin{equation}\label{inv}
  \delta n(\br_\parallel+\br_\perp;1-n_0)=-\delta n(-\br_\parallel+\br_\perp;n_0),
\end{equation}
$\br_\parallel$ and $\br_\perp$ are longitudinal and transverse components of $\br$, with respect to the external field $\bg$.

\smallskip
For relatively small obstacle with size $R$, the asymptotic behavior of the gas density perturbation $\delta n(\br)$ at $|\br|\gg R$ as well as the trace inversion effect can be obtained in the linear approximation, see \cite{kliushnychenko_blockade_2014}. Far from the obstacle, one may assume that the distribution $n=n_0+\delta n$ should weakly differ from the equilibrium one $n_0$. Thus, it is possible to linearize Eq.~(\ref{meq}) by neglecting term with $(\delta n)^2$. For the case of two-dimensional lattice gas the trace asymptotic $\delta n(\br)$ for a single obstacle of radius $R$  is given by (see \cite{kliushnychenko_blockade_2014})
\begin{eqnarray}
  \delta n(\br) \sim e^{-qr(1-\beta\cos\vartheta)}\times\quad\quad\quad\nonumber\\
  \times\sum_{n=0}\alpha_n\left[(qr)^{-\frac{1}{2}}+
  \frac{4n^2-1}{8}(qr)^{-\frac{3}{2}}+\cdots\right]\cos n\vartheta,
\end{eqnarray}
where $\vartheta$ is the angle between $\bg$ and $\br$, $r=|\br|$; constants $\alpha_n=\alpha_n(R)$ are determined from boundary conditions on the obstacle surface, $\beta=(1/2-n_0)/|1/2-n_0|=\pm1$, and $q=|1/2-n_0||\bg|$.

In the particular case of $n_0=1/2$, $\delta n$ is an odd function of $x$, see (\ref{inv}), and the density perturbation  becomes asymmetrical and extended in the $\pm x$ directions, see Fig.~\ref{fig1}b. As was shown in \cite{kliushnychenko_blockade_2014}, its asymptotic behavior damps according to the power law $\delta n(\br) \sim -(\bg, \br)R/ r^2$ that is similar to the scattered electrostatic potential for a dielectric particle in a uniform electric field.

The linear flow approximation gives the qualitative description of the trace inversion effect and the character of the asymptotic behavior of $\delta n$. In the particular case of the point-like inclusion the method gives $\delta n \sim r^{-3/2}$ that is in satisfactory agreement with numerical results and coincides with asymptotic behavior of the relaxation of the trace induced by a moving intruder presented in \cite{benichou_stokes_2000}.

However, as was mentioned above, non-linear effects may be significant near the obstacle surface.
Fig.~\ref{fig2}
shows the role of this non-linear effect in the formation of the density perturbation profile $\delta n$.
The behavior $\delta n$ near the surface has a pronounced step-like character that describes a localized dense region ahead of the obstacle at $n_0<1/2$ or a localized depleted region behind it (resembling the form of a cavity) at $n_0>1/2$.

Intuitively, such behavior can be expected. In the particular case of a lattice gas with no obstacle Eq.~(\ref{meq}), with $v\equiv 0$, admits the kink-like solution.

\section{Non-equilibrium force in a lattice gas model}

In the non-equilibrium case, there are various ways to introduce dissipative force or interaction between inclusions via Brownian gas environment. The ways are not equivalent to each other and may lead to different results in general, see \cite{Sasa_2006}. In particular, there is the Smoluchowski approach, the approaches based on a master equation, free energy, etc., see \cite{dzubiella_depletion_2003, bitbol_forces_2011,Sasa_2006,likos_effective_2001}. In this paper, we use the approach based on a master equation, see \cite{Sasa_2006}, that is close to  the often exploited method of fast variable elimination,  see, e.g., \cite{carmichael_1993,kampen_1985,nayfeh_1973}.
Considering inclusions as particles of a heavy gas component with given interaction potential between them and the original gas particles (the light component), the force (effective interaction potential) between the inclusions can  be obtained by averaging it over the light component.

\smallskip
In the equilibrium case, the average force acting on the $j$th inclusion centered at the lattice site $\bR_j$ can be formally written as (see \cite{Sasa_2006} for details)
\begin{eqnarray}\label{eforce}
  \bfF_j^{\mathit{eq}}=-\left\langle\sum_in_i\nabla_{\bR_j}V(\br_i-\bR_j)\right\rangle\nonumber\\
  = -\sum_i\langle n_i\rangle\nabla_{\bR_j}V(\br_i-\bR_j)
\end{eqnarray}
where $n_i=0, 1$ is the occupation number of site $\br_i$. The $j$th inclusion is modeled by the interaction potential $V$ between its center and the gas particles.
\begin{equation}
  \langle n_i\rangle=\sum_{\{n_i\}}\left\{n_i\rho(\{n\},0)\right\}
\end{equation}
is the averaged occupation number (gas concentration) of the $i$th site and $\rho(\{n\},0)$ is the equilibrium probability, or statistical operator in matrix representation \cite{chumak_diffusion_1980}, of finding a given occupancy configuration $\{n\}$
\begin{equation}\label{distrfunc}
  \rho(\{n\},0)=Z^{-1}\exp(-H\{n\}/kT),
\end{equation}
$Z=\sum_{\{n\}}\exp(-H\{n\}/kT)$, where $H\{n\}$ is the lattice gas-inclusions Hamiltonian
\begin{equation}
  H\{n\}=H_0\{n\}+\sum_{i,j}V(\br_i-\bR_j)n_i,
\end{equation}
and $H_0\{n\}$ is the lattice gas Hamiltonian without inclusions.

In the continuum limit and for inclusions given by a hard-core-type potential, the expression for the average force takes the custom form (see, e.g., \cite{bitbol_forces_2011})
\begin{equation}\label{force}
  \bfF_j^{\mathit{eq}}=-\int_{S_j}\bn(\br)\langle n(\br)\rangle\,\id \br,
\end{equation}
where $S_j$ is the $j$th inclusion surface and $\bn(\br)$ is its exterior normal at the point $\br$. Expression (\ref{eforce}) gives standard form for the equilibrium depletion force that can be expressed in terms of  the free energy of a lattice gas with the volume occupied by impurities being excluded \cite{Sasa_2006,dzubiella_depletion_2003}.

\medskip
In the non-equilibrium case the dissipative or non-equilibrium force can be written in a form similar to that of (\ref{eforce}), see \cite{dzubiella_depletion_2003,Sasa_2006}
\begin{eqnarray}\label{6}
  \bfF_{j}(t)=-\sum_i\left[\nabla_{\bR_j}V(\br_i-\bR_j)\right]\delta\langle n_i\rangle_t\nonumber\\
  \rightarrow -\int_S\bn(\br)\delta\langle n(\br)\rangle_t\,\id\br,
\end{eqnarray}
where $\delta\langle n_i\rangle_t=\langle n_i\rangle_t-\langle n_i\rangle$ is the density perturbation of the lattice gas and
\begin{equation}
  \langle n_i\rangle_t=\sum_{\{n\}}\left\{n_i\rho(\{n\},t)\right\}.
\end{equation}
$\rho(\{n\},t)$ is the probability of finding a given occupancy configuration $\{n\}$ at the time $t$, which satisfies the master equation for hopping process (see, e.g., \cite{gouyet_descr_2003}) with initial conditions corresponding to the equilibrium distribution (\ref{distrfunc}).

The total dissipative force, Eq.~(\ref{6}), exerted on an inclusion is defined by the induced density perturbation (obstacle trace) and can be conventionally represented by two parts. The first part is individual friction force on the isolated inclusion. The second part takes into account collective, or coherent, contribution that is related to the dissipative interaction between the obstacles, caused by their mutual influence via their traces.

\smallskip
In what follows, we will be interested in the case of steady-state non-equilibrium interaction, i.e., the limiting case $t\rightarrow\infty$.

To describe the time behavior of averaged occupation number $\langle n_i\rangle_t$ we exploit an
approach based on the kinetic equation for dynamical variable $n_i$, Eq.(\ref{balance+}),
and use the results for the average occupation number $n$ obtained in the previous section.

\section{Force switching}
\begin{figure*}
\includegraphics[width=2.05\columnwidth]{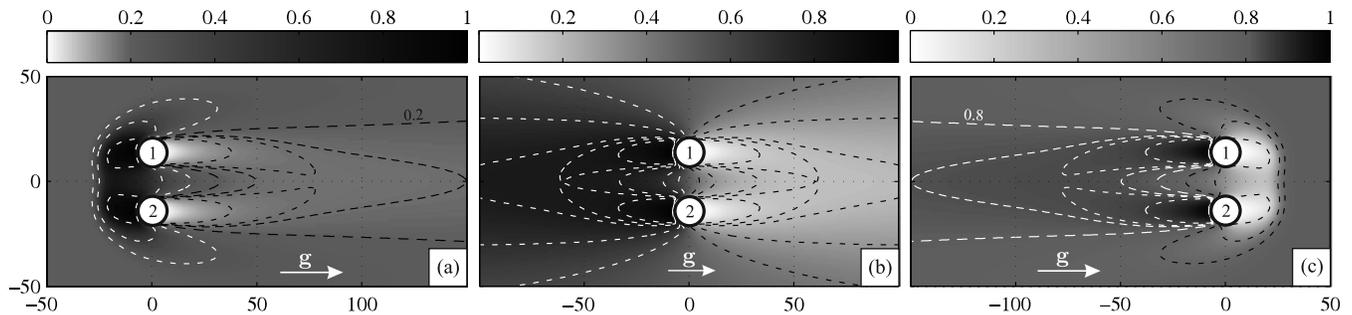}
\caption{\label{bconf} \textit{Transverse alignment.} The particle concentration distributions at three different regimes of the dissipative interaction: (a) $n_0=0.2$ \---- effective repulsion ($F^y_{12}>0$, $F^y_{21}<0$, $|F^y_{21}|=|F^y_{12}|$), (b) $n_0=0.5$ \---- no interaction ($F^y_{21}=F^y_{12}=0$), (c) $n_0=0.8$ \---- effective attraction ($F^y_{21}>F^y_{12}$). The external field $\bg$ ($|\bg|=0.5$) is directed along the $x$-axis; the impermeable ($n_0=1$) circular inclusions are of the radius $R=7$ (in units of $a$); the distance between inclusion centers $d=2y_0=4R$. The gray background corresponds to the equilibrium gas concentration $n_0$ for every contour plot, in consistence with the colorbars.}
\end{figure*}

In this section we will show the  change of sign of dissipative interaction between two obstacles, that is caused by their trace inversion.
First, we qualitatively show that such interaction switching with increasing gas concentration $n_0$  is a direct consequence of the system symmetry (\ref{inv}). To demonstrate this property we resort to the case of two inclusions located perpendicular to the gas flow, see Fig.~\ref{bconf}.
For simplicity, the inclusions are given by smooth distributions with compact carrier $v_j (\br)=v(\br-\bR_j)$, the inclusions' centers are located at points $\bR_1=(0, y_0)$ and $\bR_2=(0,- y_0)$. In this case, the force exerted on the $j$th inclusion by gas perturbation $\delta n(\br)$ reads $\bfF_j=\int\delta n\nabla v_j\id \br$. To show the possibility of the force sign changing, it is enough to consider only the $y$-component of the force  $F_j^y(n_0)$ exerted on an inclusion at equilibrium gas concentration $n_0$. Taking into account (\ref{inv}) one can obtain
\begin{eqnarray}\label{sign}
   F_j^y(n_0)=\int\delta
   n(\br;n_0)\partial_y v_j(\br)\id\br
   =-F_j^y(1-n_0).
\end{eqnarray}
In the particular case of half-filled medium ($n_0=1/2$) $F_j^y(1/2)=-F_j^y(1/2)\equiv0$, i.e., the effective interaction between the obstacles vanishes.

\smallskip
We next consider numerically the wake-mediated force between two obstacles for two configurations: the obstacles are placed parallel and perpendicular to the gas flow. The total force exerted on a given obstacle includes the part associated with individual friction force and that associated with the influence of another obstacle. To separate out the inter-obstacle contribution from the total dissipative force we consider the quantity \cite{dzubiella_depletion_2003}
\begin{equation}\label{df}
  \bfF_{ij}=\bfF_i-\bfF_i^0=\int\left[\delta n(\br,\bR_i,\bR_j)-\delta n(\br,\bR_i)\right]\nabla v_i(\br)\id\br,
\end{equation}
where $\bfF_i$ is the total force acting on the $i$th obstacle in the presence of $j$th one and $\bfF_i^0$ is its individual friction force.

\begin{figure}
\includegraphics[width=.92\columnwidth]{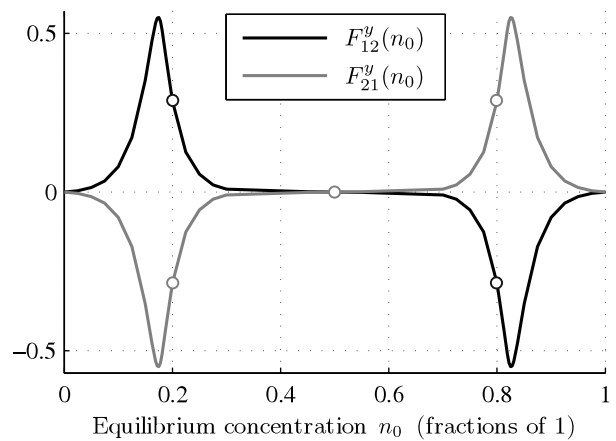}
\caption{\label{dfswitchpconfig}
The dissipative forces $F^y_{12}(n_0)$ and $F^y_{21}(n_0)$ as a functions of the equilibrium medium concentration $n_0$ in the case of transverse alignment. The values marked with circles correspond to the profiles presented on Fig.~\ref{bconf}.}
\end{figure}

\medskip
\textit{Transverse alignment} (Fig.~\ref{bconf}).
One can see from the symmetry of this configuration that the forces two obstacles exert on each other are equal and opposite, $F_{12}^y=-F_{21}^y$, i.e., their interaction is always Newtonian, see Fig.~\ref{dfswitchpconfig}.

\begin{figure}
\includegraphics[width=.98\columnwidth]{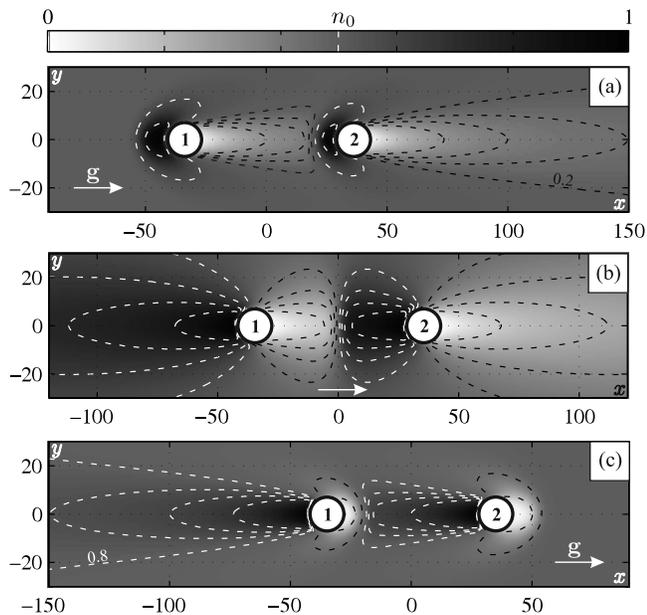}
\caption{\label{aconf} \textit{Longitudinal configuration.} The lattice gas particle concentration distributions $n(x,y)$ corresponding to three different regimes of the dissipative interaction: (a) $n_0=0.2$ \---- effective attraction ($|F^x_{21}|>|F^x_{12}|$), (b) $n_0=0.5$ \---- anti-Newtonian interaction ($F^x_{21}=F^x_{12}$), (c) $n_0=0.8$ \---- effective repulsion ($|F^x_{21}|<|F^x_{12}|$). The external field $\bg$ ($|\bg|=0.5$) is directed along the $x$-axis; the impermeable ($n_0=1$) circular inclusions are of the radius $R=7$ (in units of $a$), their positions being marked with black circles; the distance between inclusion centers $d=2x_0=10R$. The gray background corresponds to the equilibrium gas concentration $n_0$ for every contour plot, in consistence with the colorbar.}
\end{figure}
At low equilibrium concentrations ($n_0<1/2$), Fig.~\ref{bconf}a, the dissipative interaction manifests itself as an effective repulsion between the obstacles, since $F^y_{21}<0$ and $F^y_{12}>0$, see Fig.~\ref{dfswitchpconfig}. Qualitatively, this repulsion is simply explained by the overlap of density coats around the obstacles that leads to the onset of a dense region between them acting like a \textit{repelling barrier}, see Fig.~\ref{bconf}a.

In contrast, at $n_0>1/2$, the overlap of the individual density perturbation coats of the obstacles results in the formation of the extended dense zone ahead of them that blocks the gas flow, so that the region between the obstacles becomes depleted. As Fig.~\ref{dfswitchpconfig} suggests, this collective blockade effect of gas particles leads to the effective attraction between obstacles in a dense medium, $F^y_{21}>0$ and $F^y_{12}<0$.

Thus, when the gas concentration $n_0$ increases the dissipative interaction between the obstacles switches from repulsion  to attraction. In addition, the non-linear inter-obstacle attraction may characterize   the pairing effect accompanied by the creation of common perturbation coat around the obstacles. The effect of a similar nature was obtained earlier in \cite{benichou_biased_2013} for two driven intruders.

In the $n_0=1/2$ case the effective interaction between the inclusions vanishes, $F^y_{12}=F^y_{21}=0$, irrespective of the inter-inclusion distance.

\begin{figure}
\includegraphics[width=.8\columnwidth]{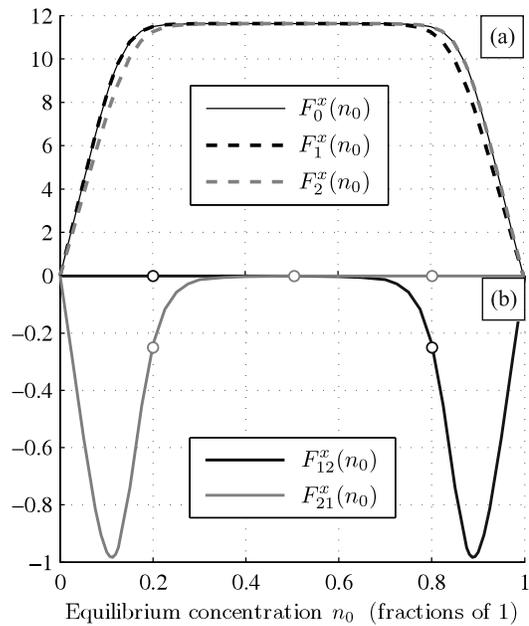}
\caption{\label{fswitchaconf}
(a) Concentration dependence of the dissipative forces $F^x_0(n_0)$, $F^x_1(n_0)$, and $F^x_2(n_0)$.
(b) The forces $F^x_{12}(n_0)$ and $F^x_{21}(n_0)$, acting between the inclusions in the longitudinal configuration. The values marked with circles correspond to the profiles presented on Fig.~\ref{aconf}.}
\end{figure}

\medskip
\textit{Longitudinal alignment} (Fig.~\ref{aconf}).
In the case of low concentration ($n_0<1/2$), the typical situation for Brownian systems is shown. An inclusion falling on the depleted wake induced by another inclusion is effectively attracted to it since the friction force in the more depleted region is weaker \cite{dzubiella_depletion_2003,khair_motion_2007}. This type of effective interaction is referred to as the wake-mediated \cite{cividini_wake-mediated_2013}. As  Fig.~\ref{fswitchaconf}b suggests, the second obstacle does not practically affects the first one, $F^x_{12}\approx0$.
In contrast, at high concentrations $n_0>1/2$, the second obstacle does not feel the influence of the first one, $F^x_{21}\approx0$, whereas the first obstacle comes under the excess pressure of the dense gas region created ahead of the second one due to the blockade effect. As a result, the effective interaction changes its sign switching from attraction to repulsion, Fig.~\ref{fswitchaconf}b.
In particular case of $n_0=1/2$, the effective interaction between the inclusions becomes strictly anti-Newtonian, $F^x_{12}=F^x_{21}\not=0$ (see Fig.~\ref{fswitchaconf}b), which holds for arbitrary inter-inclusion distance. 

Note that for a dense gas in the blockade regime, the second obstacle ``pushes'' the first one upstream, thus reducing the total friction force $F^x_1$ exerted on the first obstacle, Fig.~\ref{fswitchaconf}a.

The dissipative interaction between the inclusions naturally vanishes in the limit of strongly sparse (empty) medium $n_0\rightarrow0$, due to wake depletion. The same is true in the total jamming limit $n_0\rightarrow1$.

\section{Conclusions}\label{conclusion}

In this paper we have shown the switching of the non-equilibrium depletion force between fixed impurities or obstacles immersed in a flowing and interacting Brownian gas. To this end we have considered the simplest model of a lattice gas with two obstacles whose sizes are much larger than the lattice constant. We also used both the mean-field and the long-wavelength approximations, neglecting short-range correlations and fluctuations in the gas.

The non-equilibrium interaction between the obstacles is caused by the density gas perturbation or wakes induced around them by the gas flow.
The force switching is due to the blockade effect of the gas particles and manifests itself by changing the direction of force to its opposite.  With increasing the equilibrium gas concentration, the blockade effect provokes the wake inversion (or the wake profile turn-over) that, in turn, leads to switching of non-equilibrium depletion interaction, e.g., from the effective repulsion to attraction.

In contrast to the equilibrium case, the non-equilibrium depletion force exhibits long-range character due to extended wake tail behind the obstacles which damps according to a power law  which can be estimated in the linear approximation. However, non-linear effects become significant near obstacle. The density perturbation near its surface exhibits a pronounced step-like profile that describes the formation of a sparse cavity-like region of the gas behind or a dense drop-like region ahead the obstacle in the cases of high and low equilibrium gas concentration, correspondingly. In turn, this can lead to the effect of dissipative pairing for two closely located obstacles that accompanied by the creation of a common coat of gas density perturbation around them. However, with using the mean-field approximation we lose information on short-range correlations which may be significant near the obstacle surfaces and in the case of closely located obstacles, see \cite{tahir-kheli_correlated_1983,mejia-monasterio_bias-_2011}. Moreover, neglecting fluctuations in a gas, e.g., the  term $\delta J_i^\alpha$ in Eq.~(\ref{balance+}), we do not take into account the fluctuation-induced (Casimir-like) forces, see, e.g., \cite{bitbol_forces_2011,demery_thermal_2011,bartolo_fluctuations_2002,dean_out--equilibrium_2010,krech_fluctuation_1999,Buzzaccaro_Critical_2010,Piazza_Critical_2011}.

\acknowledgments{
We are grateful to A.A.~Chumak, B.I.~Lev, and V.V.~Gozhenko for helpful discussions and comments on the manuscript.}

\end{document}